\RequirePackage{lineno}% revtex4 users need this command BEFORE
\documentclass[11pt,a4paper]{article}
\usepackage{epsf}
\usepackage{amsmath,amssymb}
\usepackage{url,cite,ifpdf,slashed,multirow,tabularx}
\usepackage{subfigure}    % \UTF{00E5}\UTF{009B}\UTF{00B3}\UTF{00E3}\UTF{0081}\UTF{00AE}\UTF{00E4}\UTF{017E}\UTF{00AD}\UTF{00E3}\UTF{0081}\UTF{00C2}§\UTF{00E3}\UTF{0080}\UTF{0081}(a) (b) (c) \UTF{00E3}\UTF{0081}\UTF{0161}\UTF{00E3}\UTF{0081}\UTF{20AC}\UTF{00E3}\UTF{0081}\UTF{0091}\UTF{00E3}\UTF{0081}\UTF{009F}\UTF{00E3}\UTF{0081}\UTF{0084}\UTF{00E3}\UTF{0081}\UTF{0161}\UTF{00E3}\UTF{0081}\UTF{008D}\UTF{00E3}\UTF{0080}\UTF{0082}%
\usepackage{color}

\ifpdf      
\usepackage[dvipdfmx]{graphicx}                                 % For pdfLaTeX
\else                                      % For (p)LaTeX + "DVIPS"
  \usepackage[dvips]{graphicx}             %   usepackage with driver option
\fi

\newcommand{\lsim}{\mathrel{\hbox{\rlap{\lower.55ex \hbox{$\sim$}} \kern-.3em \raise.4ex \hbox{$<$}}}}
\newcommand{\gsim}{\mathrel{\hbox{\rlap{\lower.55ex \hbox{$\sim$}} \kern-.3em \raise.4ex \hbox{$>$}}}}

\def\slashchar#1{\setbox0=\hbox{$#1$} % set a box for #1
\dimen0=\wd0 % and get its size
\setbox1=\hbox{/} \dimen1=\wd1 % get size of /
\ifdim\dimen0>\dimen1 % #1 is bigger
\rlap{\hbox to \dimen0{\hfil/\hfil}} % so center / in box
#1 % and print #1
\else % / is bigger
\rlap{\hbox to \dimen1{\hfil$#1$\hfil}} % so center #1
/ % and print /
\fi}

\begin{document}

\renewcommand{\thefootnote}{\fnsymbol{footnote}} % For titlepage
\begin{titlepage}

\begin{center}

\hfill IPMU--14--0309

\vskip .75in

{\LARGE \bf Effective Interaction of Electroweak-Interacting Dark Matter with Higgs Boson and Its Phenomenology 
}

\vskip .6in

{\Large
  \textbf{Junji Hisano}$^{\rm (a,b)}$\footnote{Electronic address: hisano@eken.phys.nagoya-u.ac.jp},
  \textbf{Daiki Kobayashi}$^{\rm (a)}$\footnote{Electronic address: koba@th.phys.nagoya-u.ac.jp}, \\
  \textbf{Naoya Mori}$^{\rm (a)}$\footnote{Electronic address: m-naoya@eken.phys.nagoya-u.ac.jp}, 
  \textbf{Eibun Senaha}$^{\rm (a)}$\footnote{Electronic address: senaha@eken.phys.nagoya-u.ac.jp}, \\
}
\vskip 0.3in

$^{\rm (a)}${\em Department of Physics, Nagoya University, Nagoya 464-8602, Japan}
\vskip 0.1in
$^{\rm (b)}${\em Kavli IPMU (WPI), University of Tokyo, Kashiwa, Chiba 277--8583, Japan}

\end{center}

\vskip .6in

\begin{abstract}
  We study phenomenology of electroweak-interacting fermionic dark
  matter (DM) with a mass of $\mathcal{O}(100)$ GeV.  Constructing the
  effective Lagrangian that describes the interactions between the
  Higgs boson and the SU(2)$_L$ isospin multiplet fermion, we evaluate
  the electric dipole moment (EDM) of electron, the signal strength of
  Higgs boson decay to two photons and the spin-independent
  elastic-scattering cross section with proton.  As representative
  cases, we consider the SU(2)$_L$ triplet fermions with zero/nonzero
  hypercharges and SU(2)$_L$ doublet fermion. It is found that the
  electron EDM gives stringent constraints on those model parameter
  spaces. In the cases of the triplet fermion with zero hypercharge
  and the doublet fermion, the Higgs signal strength does not deviate
  from the standard model prediction by more than a few \% once the
  current DM direct detection constraint is taken into account, even
  if the CP violation is suppressed.  On the contrary,
  $\mathcal{O}(10$-$20)$\% deviation may occur in the case of the
  triplet fermion with nonzero hypercharge.  Our representative
  scenarios may be tested by the future experiments.

\end{abstract}

\end{titlepage}

\setcounter{page}{1}
\renewcommand{\thefootnote}{\#\arabic{footnote}}
\setcounter{footnote}{0}

%\linenumbers % to show the line numbers (need \RequirePackage{lineno})

\section{Introduction}
Nature of the dark matter (DM) in the Universe is one of the
longstanding problems in both particle physics and cosmology. The DM
abundance observed today is~\cite{Agashe:2014kda}
\begin{align}
\Omega_{\rm CDM} h^2 = 0.1198 \pm 0.0026~,
\label{omega}
\end{align}
where $h$ denotes the reduced Hubble constant. Much attention has been
paid to Weakly-Interacting Massive particles (WIMPs) as the candidates
for the DM since it is the natural consequence of physics at the TeV
scale where the next physics threshold is expected to show up based on
the naturalness argument.

Among the various DM scenarios, one of the simplest ones is that the
DM particles are coupled to the standard model (SM) particles only
through ${\rm SU(2)}_L\times {\rm U(1)}_Y$ gauge interactions. (For
earlier studies, see, {\it e.g.},
\cite{EWIMP,Cirelli:2005uq,Cirelli:2007xd}). In those cases, it is
known that the DM particle mass would be completely fixed if the
thermal relic explains the DM abundance in Eq.~(\ref{omega}).  For
example, the mass of a fermionic DM that belongs to the SU(2)$_L$
doublet (triplet) with hypercharge $Y=0$ should be about 1 (3)
TeV. Those dark matter particles are realized in the supersymmetric
(SUSY) standard model as Higgsino (Wino). In the non-thermal relic
scenarios, on the other hand, the DM relic abundance could be
satisfied as a result of a non-thermal production of the DM from late
decay of some heavy particles such as gravitinos in SUSY models. In
such a case, the DM particles do not necessarily have the multi-TeV
scale mass, and they could be as light as $\mathcal{O}(100)$ GeV.  If
so, in addition to the standard DM searches, we may find DM signals
indirectly in the collider or low energy experiments, even if they are
not directly found.

In this Letter, we study the electroweak-interacting fermionic DM
particles with the mass of $\mathcal{O}(100)$ GeV, and discuss their
phenomenological consequences in a bottom-up approach.  The
interactions between the SU(2)$_L$ isospin multiplets and the Higgs
boson are described by dimension-five operators. Such effective
interactions violate CP symmetry generically, and thus CP-violating
observables such as the electric dipole moments (EDMs) of electron,
neutron and atoms are predicted. In addition, the effective
interactions induce the spin-independent (SI) DM-nucleon
elastic-scattering cross section and the Higgs boson decay to
diphoton. In this paper we evaluate the electron EDM, the SI
DM-nucleon cross section and the Higgs signal strength for the Higgs
boson decay to diphoton mode, and confront them with current
experimental data.  Future prospects are also discussed.

\section{Models}

In this section, we will describe the effective couplings of the
fermionic DM particles with the Higgs boson. Now we assume that the DM
particle $\chi_0$ is a fermion with the SU(2)$_L\times$U(1)$_Y$ gauge
charges. The effective Higgs couplings depend on whether $\chi_0$ has
the U(1)$_Y$ interaction.

First, let us consider the case that $\chi_0$ does not have the
U(1)$_Y$ interaction ($Y=0$). In this case $\chi_0$ is a neutral
component of an isospin-$n$ multiplet $\chi_i$ ($i=-n,-n+1,\dots, +n$)
with $n$ integer. We assume for simplicity that $\chi_i$ are chiral
fermions ($\chi_i=P_L\chi_i$). The gauge interactions and the
gauge-invariant mass term are
\begin{align}
{\cal L} &= \bar{\chi} i\slashchar{D} \chi-\frac12 M (\bar{\chi^c}\chi+{\rm h.c.})~,
\label{y=0}
\end{align}
where $D_\mu = \partial_\mu+i g/\sqrt{2} (T_+ W_\mu^\dagger +T_-
W_\mu)+ig_Z (T_3-Q s_W^2)Z_\mu +ie Q A_\mu$ with $Q=T_3+Y$. Here,
$(T_\pm)_{jk}(\equiv T_1\pm i T_2)=\sqrt{n(n+1)-k(k\pm
  1)}\delta_{j,k\pm1}$, $(T_3)_{jk}=k\delta_{jk}$, and
$\bar{\chi^c}\chi =-\sum_{i=-n}^n (-1)^{i-1} \chi_i C \chi_{-i}$. The
DM particle $\chi_0$ has the Majorana mass term while other particles
with non-zero electric charges $j(\ne 0)$ have Dirac ones.  We take
$M$ real positive in the following.

The DM particle does not have renormalizable interactions with the
SU(2)$_L$ doublet Higgs boson $H$, since it is assumed to be a
fermion. Now we take the hypercharge for the Higgs boson $1/2$. The
interactions are given with higher-dimensional ones, which are induced
though integration of heavy particles. The dimension-five operators
are
\begin{align}
{\cal L}_H
&=
-\frac{1}{2 \Lambda}
|H|^2 ~\bar{\chi^c}(1 +i \gamma_5 f)\chi
% +
% \frac{1}{2 \Lambda_2}
% H^\dagger T_a H ~ \bar{\psi^c}(1 +i \gamma_5 g_{2})T_a\psi
+{\rm h.c.}~.
\label{HDO for y=0}
\end{align}
Here, only the isoscalar couplings appear at the dimension five.
While bilinears of isospin-$n$ multiplets include an SU(2)$_L$ adjoint
representation, it is antisymmetric if $n$ integer.  Those effective
interactions are induced, for example, by integration of SU(2)$_L$
$(n\pm 1/2)$-multiplet heavy fermions with hypercharge $Y=\pm 1/2$ at
the tree level. In the Wino case, the effective interaction with Higgs
boson is generated by integration of the Higgsinos.  In this paper, we
do not adopt such concrete UV models and we take a bottom-up approach
as mentioned above.

The effective couplings with the Higgs boson contribute to the masses
for $\chi_i$ after the Higgs field gets the vacuum expectation value
$(H=(0,v)^T)$ as
\begin{align}
M_{\rm phys}^2&=M_R^2+M_I^2~,
\end{align}
where
\begin{align}
M_R = M+\frac{v^2}{\Lambda}~,\quad M_I = f \frac{v^2}{\Lambda}~.
\end{align}
The masses for $\chi_i$ are degenerate at the tree level. However, it
is known that the electroweak corrections make their masses different
so that $\chi_0$ is the lightest. The mass difference between $\chi_j$
and $\chi_{j-1}$, $\Delta M_{j,j-1}$, is\footnote {The mass difference
  for $n=1$ is derived in Ref.~\cite{Cheng:1998hc}.}
\begin{align}
\Delta M_{j,j-1}&= 
\frac{\alpha_2}{4\pi}(2j-1) 
\left(f\left(x_W\right)
-c_W^2 f\left(x_Z\right)
-s_W^2 f\left(0\right)
\right) M_{\rm phys}~,
\end{align}
where
\begin{align}
f(z)&=\int^1_0 dx (2x+2) \log(x^2+(1-x) z)~.
\end{align}
Here, $x_W=m_W^2/M_{\rm phys}^2$ and $x_Z=m_Z^2/M_{\rm phys}^2$, and
$\alpha_2$ is for the SU(2)$_L$ gauge coupling constant, and
$s_W(=\sin\theta_W)$ and $c_W(=\cos\theta_W)$ are for the Weinberg
angle $\theta_W$.  When $200~{\rm GeV}\lesssim M_{\rm phys}\lesssim
3000$~GeV, $\Delta M_{j,j-1}\simeq (2j-1) \times (167$-$174)$~MeV.

Next, we present the case $\chi_0$ has the U(1)$_Y$ interaction. In
this case, $\chi_0$ comes from Dirac fermions of an isospin-$n$
multiplet, $\psi_i$ ($i=-n,-n+1,\dots, +n$).  The gauge interactions
and the gauge-invariant mass term are
\begin{align}
{\cal L} &= \bar{\psi} i\slashchar{D} \psi- M \bar{\psi}\psi~,
\label{y=nonzero}
\end{align}
and the effective interactions of $\psi$ and the Higgs boson are given
up to dimension five as
\begin{align}
{\cal L}_H=-\frac{1}{\Lambda_1} |H|^2~\bar{\psi}(1 +i \gamma_5 f_1)\psi
-\frac{1}{\Lambda_2} H^\dagger T_a H~ \bar{\psi}(1 +i \gamma_5 f_2) T_a\psi~.
\label{nonzeroY}
\end{align}
In this case, the isovector couplings are also allowed.  The physical
masses for $\psi_i$ receive the corrections from the effective
interaction after the electroweak symmetry breaking as
\begin{align}
M_{\rm phys}^{(i)2} &= M_R^{(i)2}+M_I^{(i)2},
\end{align}
where
\begin{align}
M_R^{(i)} &= M+\frac{v^2}{\Lambda_1}-\frac12 (T_3)_{ii} \frac{v^2}{\Lambda_2}~,\\
M_I^{(i)} &= f_1 \frac{v^2}{\Lambda_1}-\frac12 (T_3)_{ii} f_2\frac{v^2}{\Lambda_2}~.
\end{align}

When the first term in Eq.~(\ref{nonzeroY}) gives common corrections
to the masses, the second term induces mass splitting among the
components of multiplet. We take $\Lambda_2$ real without loss of
generality and assume it positive for simplicity. The components with
larger $T_3$ are lighter if the CP-violating coupling constant $f_2$
is negligible. Thus, the lightest state is $\psi_{n}$, and we have to
take $Y=-n$ so that the lightest state is neutral. On the other hand,
if the second term in Eq.~(\ref{nonzeroY}) is negligible, the masses
are degenerate up to the radiative corrections. In the case, the mass
difference between particles with electric charges $Q$ and $Q-1$,
$\Delta M_{Q,Q-1}$, is given as
\begin{align}
\Delta M_{Q,Q-1}&=   
\frac{\alpha_2}{4\pi} (2Q-1)
\left(f\left(x_W\right)
-c_W^2 f\left(x_Z\right)
-s_W^2 f\left(0\right)
\right) M_{\rm phys}
\nonumber\\
&+
\frac{\alpha_2}{4\pi} 2Y (
f\left(x_Z\right)-
f\left(x_W\right)) M_{\rm phys}~,
\end{align}
and then $\Delta M_{Q,Q-1}\simeq (2Q-1)\times (167$-$174)~{\rm
  MeV}+Y\times (262$-$357)~{\rm MeV}$ for $200~{\rm GeV}\lesssim
M_{\rm phys}\lesssim 3000$~GeV. Thus, the neutral fermion is the
lightest only when $Y=\pm n$, unless the tree- and loop-level
contributions to the mass of the neutral fermion cancel each others
accidentally.

Null results of the DM direct detection give a stringent constraint on
the vector coupling of the dark matter particles. Thus, the DM
particle has to be a Majorana fermion in order to forbid the vector
current interaction. Now let us consider the case of $Y=-n$.  The
neutral component $\psi_n$ with $T_3=n$ is decomposed into the
Majorana fermions ($\chi_0$ and $\chi_0^\prime$) as
\begin{align}
\psi_n=\chi_0+i \chi_0^\prime~,
\end{align}
and they should not be degenerate in mass, so that the DM direct
detection is suppressed.  Such mass splitting is generated by the
following fermion-number violating interaction,
\begin{align}
{\cal L}_{HM}&= \frac1{\Lambda_M^{4n-1}}[H^{2n}\bar{\psi^c}] [H^{2n}\psi]
+{\rm h.c.}~.
%\Lambda_M \left(\prod_{i=1}^{2m}\frac{H T_{a_i} H}{\Lambda_M^2} \right)
%\bar{\psi^c} (T_{a_1}\cdots T_{a_{2m}})\psi
\end{align}
where $[\cdots]$ is SU(2)$_L\times $U(1)$_Y$ invariant.  For example,
when $n$ is integer, this interaction is generated by integration of
heavy fermions of isospin-$(i-1/2)$ multiplets with $Y=-(i-1/2)$ and
of isospin-$(i-1)$ multiplets with $Y=-(i-1)$ $(i=1,\dots, n)$, which
have the Yukawa couplings with the Higgs boson. The isospin singlet
fermion with $Y=0$ has the Majorana mass which is a source of the
fermion number violation.

For later use, we collectively denote the Higgs couplings to the
fermion fields ($\Psi=\chi_i,~\psi_i$) in the rotated basis as
\begin{align}
%\mathcal{L}_H & =  -c\bar{\chi_i}\left(g_S^{(i)}+i\gamma_5g_P^{(i)}\right)\chi_ih,
%\mathcal{L}_H & =  -c\bar{\chi}\left(g_S+i\gamma_5g_P\right)\chi h~,
\mathcal{L}_H & =  -c{\bar{\Psi}}\left(g_S+i\gamma_5g_P\right){\Psi} h~,
\end{align}
where $c=1~(1/2)$ for Dirac (Majorana) fermions and 
\begin{align}
g_S & = \frac{1}{|M_{\rm phys}|} \left(g_S'M_R+g_P'M_I\right), \\
g_P & = \frac{1}{|M_{\rm phys}|} \left(-g_S'M_I+g_P'M_R\right),
\end{align}
with $g_{S,P}'$ being the Higgs couplings to the fermion fields in the original basis.

\section{Electron EDM, DM direct detection, and Higgs to two gammas}

The axial-scalar couplings in Eqs.~(\ref{HDO for y=0}) and
(\ref{nonzeroY}) are CP violating, so that the EDMs are generated. Now
the electron EDM is severely bounded from above from the ACME
experiment~\cite{dfh:2011},
\begin{align}
|d_e|<8.7\times 10^{-29} e{\rm cm}~,
\end{align}
and the bound gives the CP-violating couplings more severe than other
EDMs. In this paper we consider constraints on the models using the
electron EDM. The bound would be improved to $10^{-(29-30)} e{\rm cm}$
in near future.

%---------------------------------------------------
\begin{figure}[t]
\center
\includegraphics[width=7cm]{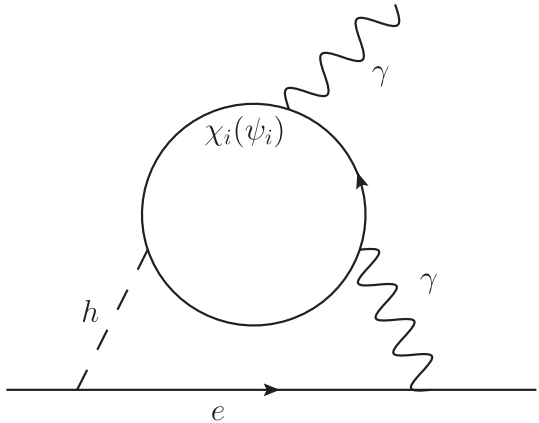}
\caption{Barr-Zee diagrams.}
\label{fig:BZ}
\end{figure}
%---------------------------------------------------

The CP-violating Higgs couplings contribute to the electron EDM via
the Barr-Zee diagrams at the two-loop level~\cite{Barr:1990vd}.  The
Barr-Zee diagrams include the CP-violating anomalous
$\gamma$-$\gamma$-$h$ coupling induced after integrating out $\chi$ in
Eq.~(\ref{HDO for y=0}) or $\psi$ in Eq.~(\ref{nonzeroY}), given in
Fig.~\ref{fig:BZ}.  The CP-violating anomalous $\gamma$-$Z$-$h$
coupling is also present, though the contribution to the electron EDM
is suppressed due to the accidentally suppressed vector coupling of
$Z$ boson with electron, $(1/4-\sin\theta_W^2)\simeq 0.02$.  The
CP-violating anomalous $\gamma$-$W^+$-$W^-$ coupling also contributes
to the electron EDM. However, the anomalous coupling should be zero at
the one-loop level if the components in isospin multiplets are
degenerate in mass. Thus, this contribution is also negligible.

By evaluating the diagrams in Fig.~\ref{fig:BZ}, we derived the
electron EDM in the model in Eq.~(\ref{HDO for y=0}) as
\begin{align}
\frac{d_e}{e}&=
\frac{\alpha}{8\pi^3}\frac{m_eM}{M_{\rm phys}^2}\frac{f}{\Lambda} A_n
F(z)~,
\end{align}
where the mass function $F(z)$ $(z=M_{\rm phys}^2/m_h^2)$ is
\begin{align}
F(z)&=\frac 12 z \int^1_0  dt \frac1{t(1-t)-z} \log\frac{t(1-t)}{z}~.
\end{align}
When $z\gg 1$, $F(z)\simeq\frac12 \log z+1$. Since $A_n$ is
\begin{align}
A_n&=\frac16 n(n+1)(2n+1)~,
\end{align}
the electron EDM is enhanced for large $n$. In this paper we use
$m_h=125.5$~GeV~\cite{Aad:2014aba,CMS:2014ega}.

The electron EDM in the model in Eq.~(\ref{nonzeroY}) is 
\begin{align}
\frac{d_e}{e}&=
\sum_i \frac{\alpha}{8\pi^3}\frac{m_eM}{M^{(i)2}_{\rm phys}}\left(\frac{f_1}
{\Lambda_1}-\frac{f_2}{2\Lambda_2}(T_3)_{ii}\right)Q_i^2
F(z_i)~,
\label{de_Yneq0}
\end{align}
where $Q_i=(T_3)_{ii}+Y$. When the masses are degenerate in the
multiplets, the electron EDM is reduced as
\begin{align}
\frac{d_e}{e}&=
\frac{\alpha}{8\pi^3}\frac{m_eM}{M^{(i)2}_{\rm phys}}\left(\frac{f_1}
{\Lambda_1}A_n^{(1)}-\frac{f_2}{2\Lambda_2} A_n^{(2)}\right)
F(z)~, 
\end{align}
where 
\begin{align}
A_n^{(1)}&=\frac{1}{3}(n(n+1)+3Y^2)(2n+1)~, \nonumber\\
A_n^{(2)}&=\frac{2}{3}n(n+1) (2n+1) Y~.
\end{align}
Then, the contribution from $f_2$ is more enhanced, especially when
$n=\pm Y$.

The Barr-Zee diagram contribution to the electron EDM is suppressed by
$M_{\rm phys}$ and $\Lambda$. We evaluate the EDM using the effective
higher-dimensional interactions in Eq.~(\ref{HDO for y=0}) or $\psi$
in Eq.~(\ref{nonzeroY}). While those interactions are generated by the
some heavy fields, the contributions from the heavy fields to the EDM
are suppressed by $\Lambda^2$. Thus, we may neglect them as far as
$M_{\rm phys}\ll \Lambda$, and the effective theory description works
well for evaluation of the EDM.\footnote{ The electron EDM is
  evaluated in UV theories for Wino and Higgsino in
  Ref.~\cite{Giudice:2005rz}.} This is also valid when we consider the
Higgs boson decay to diphoton.

In this paper we use the electron EDM bound in order to constrain the
parameter spaces. The quark EDMs induced by the Barr-Zee diagrams are
related to the electron EDM as $d_q/d_e=(Q_q/Q_e)(m_q/m_e)$. The
neutron EDM is given as $d_n\simeq 0.79d_d-0.20 d_u$ from the QCD sum
rules \cite{Hisano:2012sc} and the latest neutron EDM bound is
$|d_n|<2.9\times 10^{-26}e$cm \cite{Baker:2006ts}. Thus, the neutron
EDM bound is not as stringent as the electron one now.

Next is the Higgs boson decay to two gammas. The signal strength of
the Higgs boson decay to two gammas is determined by the low-energy
theorem \cite{Voloshin:2012tv,McKeen:2012av,Fan:2013qn}, and the
contribution from the $n$-multiplet fermions is included as
\begin{align}
\mu_{\gamma\gamma}
&=
\left|1+\frac{G_R}{A_{\rm SM}}  \right|^2
+
\left|\frac{G_I}{A_{\rm SM}} \right|^2,
\label{mugamgam}
\end{align}
where the SM amplitude $A_{\rm SM}$ is $-6.49$. Here
\begin{align}
G_R&=
\sum_i\frac{4\sqrt{2}Q_i^2v}{3|M_{\rm phys}^{(i)}|}g_S^{(i)}~,\\ 
G_I&=
\sum_i\frac{2\sqrt{2}Q_i^2v}{|M_{\rm phys}^{(i)}|}g_P^{(i)}~.
\end{align}
The measured values of $\mu_{\gamma\gamma}$ at the
ATLAS~\cite{Aad:2014eha} and CMS~\cite{Khachatryan:2014ira} are
respectively given by
\begin{align}
\mu_{\gamma\gamma}=1.17\pm0.27~{(\rm ATLAS)}~,\quad
\mu_{\gamma\gamma}=1.14_{-0.23}^{+0.26}~{(\rm CMS)}~.
\end{align}

%---------------------------------------------------
\begin{figure}[t]
\center
\includegraphics[width=10cm]{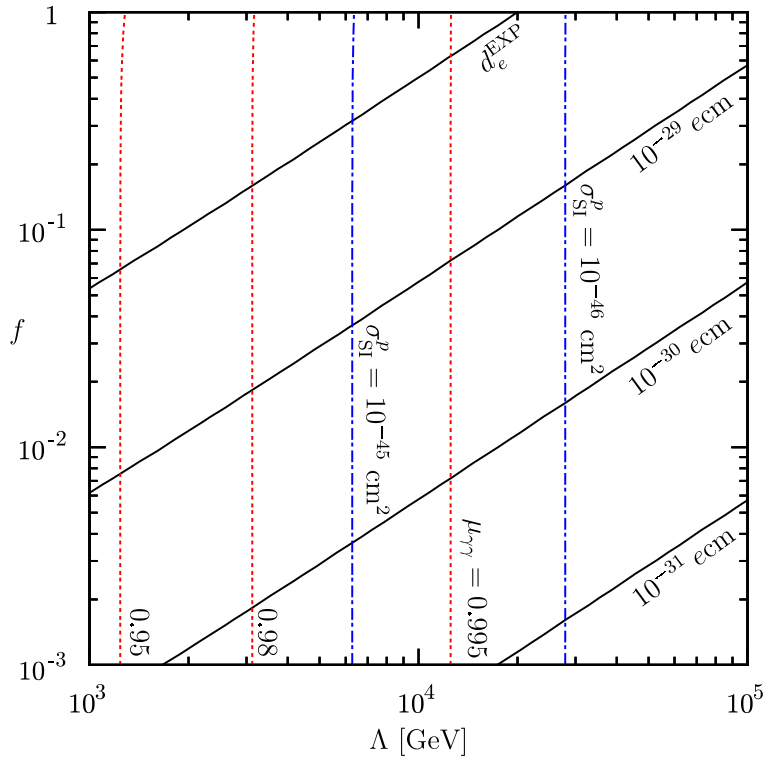}
\caption{$|d_e|$, $\sigma_{\rm SI}^p$ 
and $\mu_{\gamma\gamma}$ in the case of $Y=0$ and $n=1$ (triplet). 
We set $M_{\rm phys}=400$ GeV.}
\label{fig:Lam_f_n1Y0}
\end{figure}
%---------------------------------------------------

Last, we consider the DM direct detection by elastic scattering with
nucleon. It is induced by the DM particle coupling with the Higgs
boson.  The SI cross section of the DM particle $\chi^0$ with proton
is at the leading order,
\begin{align}
\sigma_{\rm SI}^p&=
\frac{2}{\pi}
\frac{m_p^4g_S^2}{m_h^4 v^2} \left(\frac19 f_{TG}+\frac12 \sum_{q=u,d,s}f_{ Tq}\right)^2~.
\label{sigma_tree}
\end{align}
(See references \cite{Jungman:1995df,Hisano:2010ct}.)  Here, $m_p$ is
the proton mass, and $f_{TG}$ and $f_{Tq}$ are the nucleon mass
fractions of gluon and quarks, respectively. { In this paper we use
  $f_{ Tu}=0.023$, $f_{Td}=0.032$, $f_{ Ts}=0.020$, $f_{
    TG}=1-\sum_{q=u,d,s}f_{Tq}=0.925$~\cite{Hisano:2011cs}.}  Even if
the DM particle has only the electroweak interaction, the SI cross
section with proton of the DM particle is induced at the loop
level. The contribution is evaluated in Ref.\cite{Hisano:2011cs}.
Since such contributions may be relevant in the large $\Lambda$ region
where Eq.~(\ref{sigma_tree}) is suppressed, we take them into account
in our numerical calculation.  Recently, LUX
experiment~\cite{Akerib:2013tjd} put a new constraint on $\sigma_{\rm
  SI}^p$.  For example, $\sigma_{\rm SI}^p\lesssim10^{-45}~{\rm cm}^2$
for the DM with a mass of 100 GeV.

Before going into the numerical analysis, the experimental constraints
on $M_{\rm phys}$ are discussed.  First, we consider the bounds coming
from the Large Hadron Collider (LHC).  In the case of $\Delta
M_{Q,Q-1}\lsim 1$ GeV, the dominant decay mode of $\chi_Q$ is
$\chi_Q\rightarrow \chi_{Q-1}\pi^+$. In the Wino case, which is the
typical lifetime of $\chi^\pm$ is $\mathcal{O}(0.1)$ ns.  Such a
metastable charged particle may be probed by looking at the
disappearing charged track at the LHC.  Using 20.3 ${\rm fb}^{-1}$
data collected at $\sqrt{s}=8$ TeV running, the ATLAS collaboration
places a lower bound on $M_{\rm phys}$~\cite{Aad:2013yna}
\begin{align}
  M_{\rm phys} > 270~{\rm GeV}~,
\end{align}
at 95\% C.L. For larger $n$, however, the lifetime of $\chi$ is so
short that we do not obtain a useful constraint by this searches.

When the mass splitting is larger than $\sim$ 1~GeV, the constraints
on $M_{\rm phys}$ from the LHC experiments are weaker.  The searches
for direct production of charginos and neutralinos in final states
with two or three leptons and missing transverse energy are conducted
at the LHC.  Depending on the lightest neutralino mass (denoted as
$m_{\tilde{\chi}^0}$), the limit is placed on the masses of the
chargino and the second lightest neutralino which are assumed to be
degenerate.  For example, the chargino has to be heavier than 415 GeV
if $m_{\tilde{\chi}^0}=0$ while no significant bound is obtained if
$m_{\tilde{\chi}^0}\gtrsim160$ GeV~\cite{Aad:2014nua,Aad:2014vma}.

The other constraints may come from indirect DM searches.  The
comprehensive studies on the indirect detection of the Wino DM are
conducted in Ref.~\cite{Bhattacherjee:2014dya}.  The Wino mass is
bounded as
\begin{align}
320~{\rm GeV}\lesssim M_{\rm phys} \lesssim 2.25~{\rm TeV}~,\quad 
2.43~{\rm TeV}\lesssim M_{\rm phys} \lesssim 2.9~{\rm TeV}~,
\end{align}
which are set by gamma-ray observations of classical dwarf spheroidal
galaxies and the DM relic abundance constraint. For larger $n$, the
constraints would become more severe if the DM dominates the observed
DM abundance in Eq.~(\ref{omega}), since the annihilation cross
sections of the DM grow as $\mathcal{O}(n^4)$
\cite{Cirelli:2005uq,Cirelli:2007xd}.  In this paper, we consider the
cases of $n=1$ and $n=1/2$, and the other cases are discussed
qualitatively.

\section{Results}

Now, we show our numerical results.  First, we consider a case in
which $Y=0$ and $n=1$ (isospin triplet).  In
Fig.~\ref{fig:Lam_f_n1Y0}, $|d_e|$, $\sigma_{\rm SI}^p$ and
$\mu_{\gamma\gamma}$ are plotted in the $(\Lambda, f)$ plane. We take
$M_{\rm phys}=400$ GeV as an example.
In this case, $370$ GeV
$\lesssim M \lesssim$ 400 GeV.  The black lines, from top to bottom,
represent $|d_e^{\rm EXP}|=8.7\times10^{-29}~e{\rm cm}$,
$10^{-29}~e{\rm cm}$, $10^{-30}~e{\rm cm}$ and $10^{-31}~e{\rm cm}$,
respectively.  As shown, $\Lambda$ has to be greater than around 2(20)
TeV for $f=0.1(1.0)$ to satisfy the current limit.  When $M_{\rm
  phys}$ is heaver, $|d_e|$ is scaled as $1/M_{\rm phys}$.

The dotted-dashed vertical lines in blue denote $\sigma_{\rm
  SI}^p=10^{-45}~{\rm cm}^2$ and $\sigma_{\rm SI}^p=10^{-46}~{\rm
  cm}^2$ from left to right. It is insensitive to the CP-violating
coupling $f$ and also $M_{\rm phys}$.  It is found that
Eq.~(\ref{sigma_tree}) becomes smaller than the loop contributions
($\simeq1.4\times10^{-47}~{\rm
  cm^2}$)~\cite{Hisano:2010ct,Hisano:2011cs} for $\Lambda\gtrsim 47$
TeV, and reach around $3.0\times10^{-48}~{\rm cm^2}$ at $\Lambda=100$
TeV.

The dotted lines in red represent $\mu_{\gamma\gamma}=0.95, 0.98$ and
0.995 from left to right.  The numerical impact of the fermions with
$|Q|=1$ on $\mu_{\gamma\gamma}$ is less than 2\% for $\Lambda\gtrsim
3$ TeV.  As seen in Eq.~(\ref{mugamgam}), the dominant new physics
contribution comes from the CP-conserving part. The deviation of
$\mu_{\gamma\gamma}$ from one is scaled as $1/(\Lambda M_{\rm phys})$,
and it is less sensitive to $f$.

%---------------------------------------------------
\begin{figure}[t]
\center
\includegraphics[width=10cm]{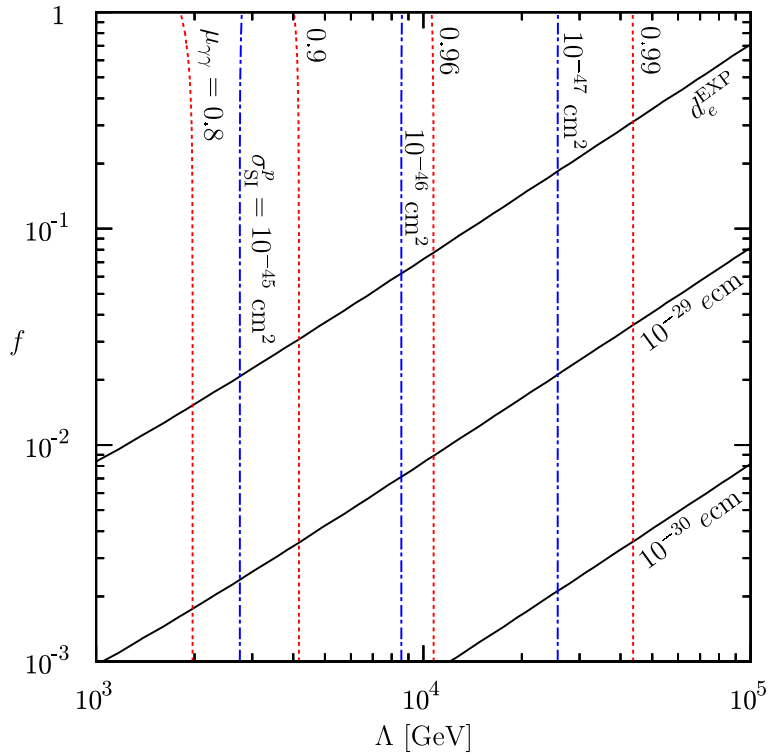}
\caption{$|d_e|$, $\sigma_{\rm SI}^p$ 
and $\mu_{\gamma\gamma}$ in the case of $n=1=-Y$ (triplet). We set $f_1=f_2\equiv f$, 
$\Lambda_1=\Lambda_2\equiv \Lambda$ and $M=400$ GeV.}
\label{fig:Lam_f_n1Yneq0}
\end{figure}
%---------------------------------------------------

Next, we illustrate a case in which $n=1=-Y$. Our finding is shown in
Fig.~\ref{fig:Lam_f_n1Yneq0}.  For simplicity, we take $f_1=f_2\equiv
f$, $\Lambda_1=\Lambda_2\equiv \Lambda$ and $M=400$ GeV.  Similar to
the previous case, we display $|d_e|$, $\sigma_{\rm SI}^p$ and
$\mu_{\gamma\gamma}$ in the $(\Lambda, f)$ plane.  The color scheme is
the same as in Fig.~\ref{fig:Lam_f_n1Y0}.  The fermion masses with
$|Q|=0, 1, 2$ are varied in the following ranges: 400 GeV $\lesssim
M_{\rm phys}^{(1)}\lesssim 415$ GeV, 400 GeV $\lesssim M_{\rm
  phys}^{(0)}\lesssim 431$ GeV and 400 GeV $\lesssim M_{\rm
  phys}^{(-1)}\lesssim 448$ GeV, respectively.  Since there are the
fermions with $|Q|=2$ in this case, the phenomenological consequences
are quite different from the previous case.  As for the electron EDM
constraint, the region where $|d_e/d_e^{\rm EXP}|\leq 1$ is satisfied
gets significantly smaller.  For instance, $\Lambda$ must be greater
than around 15 TeV for $f=0.1$, and $f=1$ is not allowed even if
$\Lambda=100$ TeV.

We observe that $\sigma_{\rm SI}^p$ is reduced to some extent compared
to the previous case.  In the large $\Lambda$ region, $\sigma_{\rm
  SI}^p$ may be as small as $\mathcal{O}(10^{-48})~{\rm cm}^2$.  As
shown in Ref.~\cite{Hisano:2011cs}, the loop contributions in
$\sigma_{\rm SI}^p$ get reduced as $Y$ increases.

It is also found that $\mu_{\gamma\gamma}$ may be significantly
reduced. For example, $\mu_{\gamma\gamma}$ may be as small as 0.8 at
$\Lambda\simeq 2$ TeV, which is mainly due to the contribution of the
fermions with $|Q|=2$.

In this example, we simply assume $f_1=f_2$ and $\Lambda_1=\Lambda_2$.
However, we could treat them separately.  In such a case, we may have
sizable cancellations in $d_e$ by choosing the input parameters
judiciously as inferred from Eq.~(\ref{de_Yneq0}), leading to more
relaxed bounds.

%---------------------------------------------------
\begin{figure}[t]
\center
\includegraphics[width=10cm]{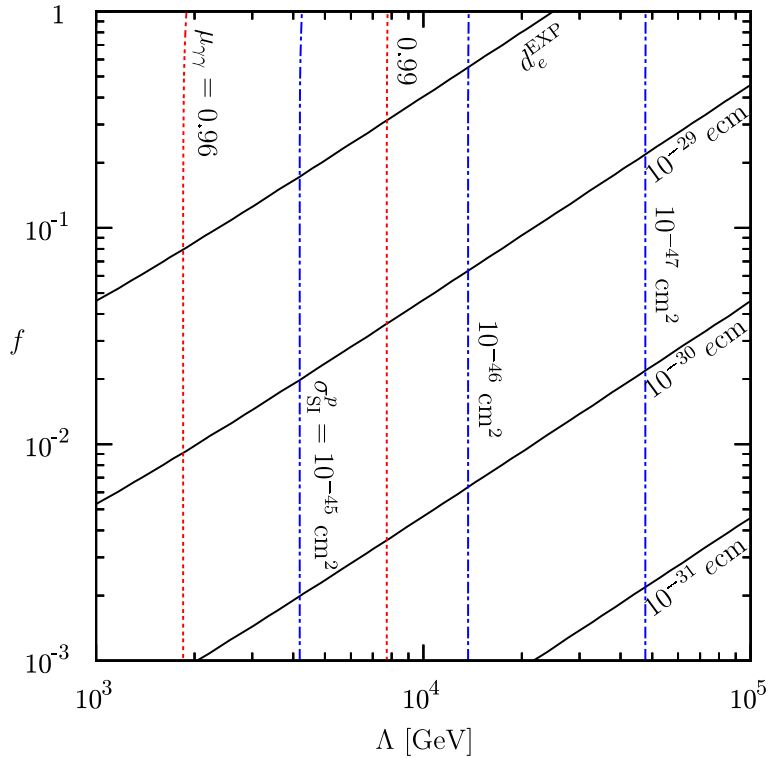}
\caption{$|d_e|$, $\sigma_{\rm SI}^p$ 
and $\mu_{\gamma\gamma}$ in the case of $n=1/2=-Y$ (doublet). We set $f_1=f_2\equiv f$, 
$\Lambda_1=\Lambda_2\equiv \Lambda$ and $M=400$ GeV.}
\label{fig:Lam_f_nhalfYneq0}
\end{figure}
%---------------------------------------------------

The results in the case of $n=1/2=-Y$ (doublet) are shown in
Fig.~\ref{fig:Lam_f_nhalfYneq0}.  The input parameters are the same as
in Fig.~\ref{fig:Lam_f_n1Yneq0}.  The masses of the fermions with
$|Q|=0,1$ are varied in the ranges: 400 GeV $\lesssim M_{\rm
  phys}^{(1/2)}\lesssim 423$ GeV, 400 GeV $\lesssim M_{\rm
  phys}^{(-1/2)}\lesssim 440$ GeV, respectively.  Although the
particle content is similar to the case of $Y=0$ and $n=1$, the
electron EDM and $\mu_{\gamma\gamma}$ are somewhat enhanced due to the
presence of the second term in Eq.~(\ref{nonzeroY}).  On the other
hand, $\sigma_{\rm SI}^p$ gets smaller because of $Y\neq0$
contributions at the loop level as mentioned above.

We considered the cases of $(n,Y)=(1,0),~(1,-1),~(1/2,-1/2)$.  When
$n$ is increased, fermions with larger electric charges are
introduced. When $Y$ is zero, the electron EDM and deviation of
$\mu_{\gamma\gamma}$ from one are scaled by $A_n=n(n+1)(2n+1)/6$. When
$Y$ is nonzero, fermions with larger electric charges than in the
cases of $Y=0$ contribute to them so that the larger effects are
expected. Furthermore, when the isovector couplings are sizable, the
contributions to $d_e$ and $\mu_{\gamma\gamma}$ dominate over those
from the isoscalar couplings.  On the other hand, $\sigma^p_{\rm SI}$
is insensitive to $n$ when $Y=0$.  If $Y\ne 0$, $\sigma^p_{\rm SI}$
depends on $n$ and $Y$ via the isovector coupling $f_2$.

%---------------------------------------------------
\begin{figure}[t]
\center
\includegraphics[width=8cm]{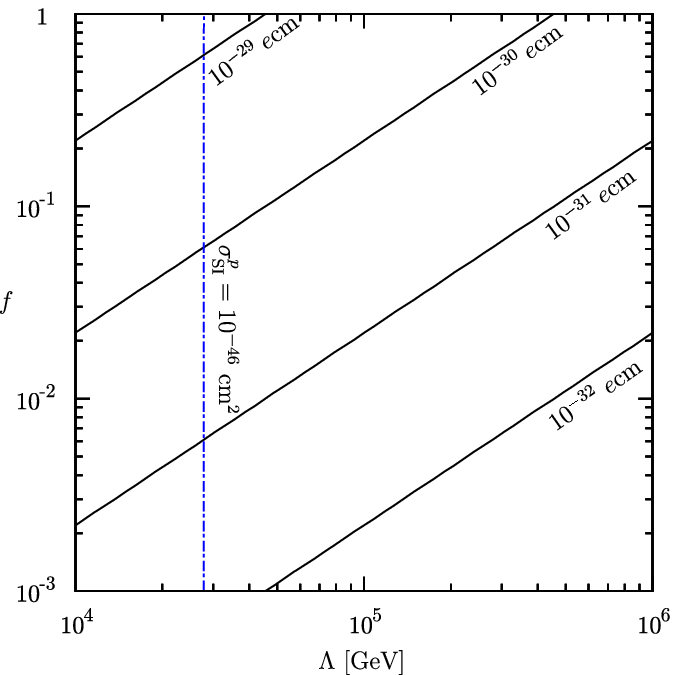}
\includegraphics[width=8cm]{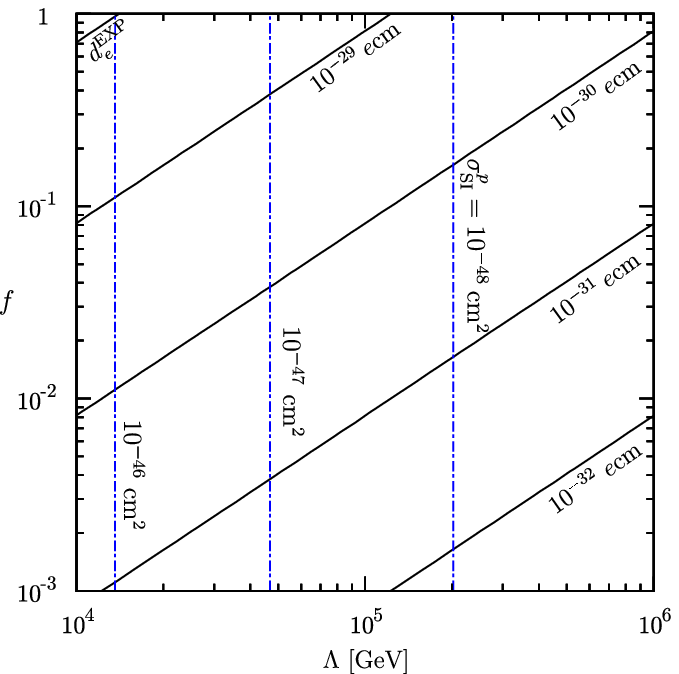}
\caption{$|d_e|$ and $\sigma_{\rm SI}^p$ in the thermal DM scenario.
(Left) $Y=0$ and $n=1$, $M_{\rm phys}=2900$ GeV; (Right) $n=1/2=-Y$, $M=1000$ GeV.}
\label{fig:Lam_f_therDM}
\end{figure}
%---------------------------------------------------
Here, we briefly discuss the thermal DM scenario. 
The results of $(n,Y)=(1,0),~(1/2,-1/2)$ are presented in Fig.~\ref{fig:Lam_f_therDM}.
We do not consider the $(n,Y)=(1,-1)$ case since the correct DM mass 
in light of the Sommerfeld effect is unknown.
Notice that $\Lambda$ is taken from $10^4$ GeV to justify our analysis based on the 
effective Lagrangians (\ref{HDO for y=0}) and (\ref{nonzeroY}). 
Since the mass scale of each multiplet is 2900 GeV and 1000 GeV, 
the deviation of $\mu_{\gamma\gamma}$ from the SM value is less than 1\% so that
the contour is not shown here.
Likewise, the current electron EDM bound is not strong enough to probe 
this parameter space except the tiny portion in the case of $(n,Y)=(1/2, -1/2)$. 
As for the DM direct detection in the $(n,Y)=(1, 0)$ case, 
$\sigma_{\rm SI}^p$ mildly decreases as $\Lambda$ increases
and reach $1.5\times 10^{-47}~{\rm cm}^2$ at $\Lambda=10^6$ GeV. 
In the $(n,Y)=(1/2, -1/2)$ case, on the other hand, $\sigma_{\rm SI}^p$ is more suppressed
due to the $Y\neq0$ contributions in the loop corrections as mentioned above. 

Finally, we remark some future prospects.  The electron EDM is
expected to be improved up to $\sim 10^{-30}~e{\rm cm}$
level~\cite{Sakemi:2011zz,Kara:2012ay,Kawall:2011zz}.  For the DM
direct detection, $\sigma_{\rm SI}^p$ would be improved by more than
one order of magnitude by the XENON1T experiment~\cite{Aprile:2012zx},
and further improvement is projected by the LZ
experiment~\cite{Cushman:2013zza}.  The future collider experiments
such as the high-luminosity LHC
(HL-LHC)~\cite{ATLAS:2013hta,CMS:2013xfa} and International Linear
Collider (ILC)~\cite{Baer:2013cma} may improve the sensitivity of
$\mu_{\gamma\gamma}$ up to $\mathcal{O}(5)$\%.  Combining all data,
especially the former two, we may test almost entire region in the nonthermal DM scenario.

\section{Conclusion}

We have studied phenomenology of electroweak-interacting fermionic
dark matter with a mass of $\mathcal{O}(100)$ GeV.  Constructing the
effective Lagrangian that describes the interactions between the Higgs
boson and the SU(2)$_L$ isospin multiplet fermion, we evaluate the
electric dipole moment of electron, the signal strength of Higgs boson
decay to two photons and the DM direct detection.

As a benchmark point, 400 GeV for SU(2)$_L$ isospin multiplet fermion
mass is taken throughout this analysis.  In particular, we
investigated the three representative cases: (1) triplet fermion with
$Y=0$, (2) triplet fermion with $Y\neq0$ and (3) doublet fermion with
$Y\neq0$.  It is found that the LUX direct detection bound
($\sigma_{\rm SI}^p\sim 10^{-45}~{\rm cm}^2$) is probing
$\Lambda\simeq$ multi-TeV region, and the case (1) suffers from it the
most among the three.  If the CP-violating Higgs-fermion coupling is
unity, the current electron EDM limit pushes the cutoff $\Lambda$ up
to around $20$ TeV in the cases (1) and (3) while above 100 TeV in the
case (2).  In light of the current DM direct detection constraint, the
signal strength of the Higgs boson decay to two photons deviates from
the SM value by more than a few \% in the cases (1) and (3).  In the
case (2), on the other hand, about 20\% deviation of Higgs signal
strength is still allowed even though it is out of 1$\sigma$ range of
the current data.  Our analysis shows that the unconstrained areas in
the all cases would be tested by the future improvements of $d_e$ and
$\sigma_{\rm SI}^p$. The HL-LHC and ILC are also important to probe
the region of $\Lambda\lesssim 10$ TeV in the case (2).

\section*{Acknowledgments}
The work of J.H. is supported by Grant-in-Aid for Scientific research
from the Ministry of Education, Science, Sports, and Culture (MEXT),
Japan, No. 24340047 and No. 23104011, and also by World Premier
International Research Center Initiative (WPI Initiative), MEXT,
Japan. The work of D.K. is supported by Grant-in-Aid for Japan Society
for the Promotion of Science (JSPS) Fellows (No. 26004521).

%%%%%%%%%%%%%%%%%%%%%%%%%%%%%%%%%%%%%%%%%%%%%
\providecommand{\href}[2]{#2}
\begingroup\raggedright

\endgroup

\begin{thebibliography}{99}
\bibitem{CMS:2014ega}
  CMS Collaboration [CMS Collaboration],
  %``Precise determination of the mass of the Higgs boson
and studies of the compatibility of its couplings with the standard model,''
  CMS-PAS-HIG-14-009.
\end{thebibliography}
\end{document}